\def\kppg  {K^{0}_{L} \longrightarrow \pi^+ \pi^- \gamma}
\def\kppgs {K^{0}_{L} \longrightarrow \pi^+ \pi^- \gamma^*}
\def\kppee {K^{0}_{L} \longrightarrow \pi^+ \pi^- e^+ e^- }
\def\kpip  {K^{0}_{L} \longrightarrow \pi^+ \pi^- }
\def\kpmz  {K^{0}_{L} \longrightarrow \pi^+ \pi^- \pi^0 }
\def\reps  {{\cal R}e ( \epsilon^{\prime} / \epsilon)}
\documentstyle[aps,epsf]{revtex}  % DO NOT DELETE THIS LINE 

\begin{document}        %  DO NOT DELETE OR CHANGE THIS LINE

\baselineskip 14pt
\title{Study of the Decays $\kppg$ and $\kppee$ at KTeV}
\author{John Belz}
\address{Rutgers University}
% \author{}   % Use this and the next line only if there is a second
% \address{Another University, etc.}  % address. (Remove the left % marks)
%
\maketitle              % Creates the title area, Do Not Remove

\begin{abstract}        % Do Not Delete this line
We present new results on the related rare $K^0_L$ decay modes  
$\kppg$ and $\kppee$. KTeV has performed 
the first direct measurement of the form factor for the 
``direct emission'' component of $\kppg$ decays, 
a quantity with ramifications for particular chiral models. 
In addition, the form factor and direct emission/inner 
bremsstrahlung branching ratio --- also presented here --- are 
important input parameters for the understanding of the 
planar--angle distribution of $\kppee$ decays. 
Preliminary results indicating the presence of a T--violating 
asymmetry in the $\kppee$ angular distribution are presented.
\
\end{abstract}   	% Do Not Delete this line

\section{Introduction}               % Introduction goes below.

In this report we discuss recent studies of the related rare 
$K^0_L$ decay modes $\kppg$ and $\kppee$ by the KTeV 
experiment at Fermilab. The decays share the ``Inner Bremsstrahlung'' 
(IB) and ``Direct Emission'' (DE) diagrams of Figure~\ref{diagrams}:
\begin{figure}[ht]	% in second brace, h=here, t=top, b=bottom	
\centerline{ \epsfxsize 1.0 truein \epsfbox{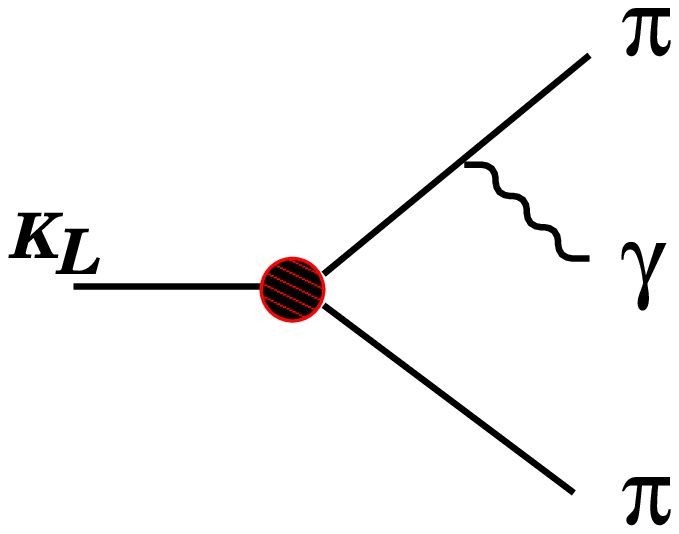} \hspace{1.0cm}  
             \epsfxsize 1.0 truein \epsfbox{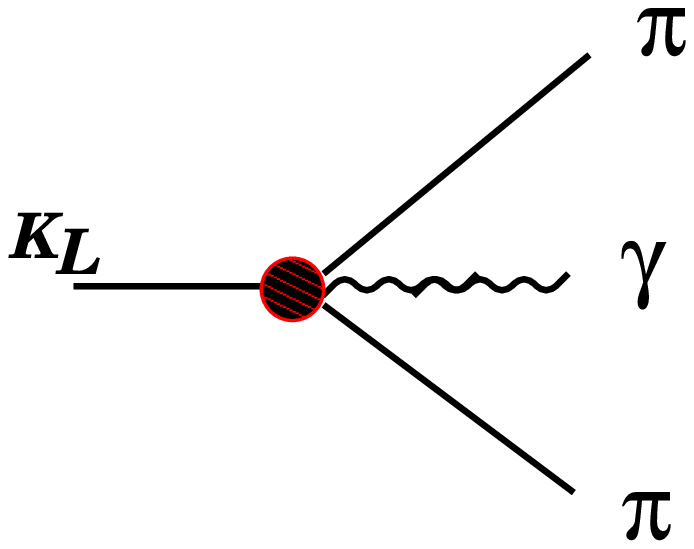} \hspace{1.0cm}
             \epsfxsize 1.0 truein \epsfbox{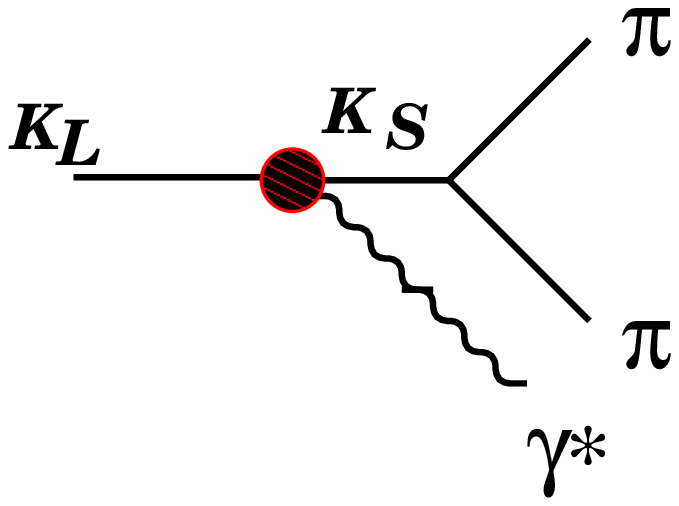} }  
\vspace{0.3cm}
\caption[]{
\label{diagrams}
\small Contributions to $\kppgs$ from (left) Inner Bremsstrahlung,
  (center) direct emission, and (right) $K^0$ charge radius.}
\end{figure}
The two decays differ in that the photon converts internally in 
the $\kppee$ case. A third diagram known as the ``$K^0$ Charge Radius'' 
contribution contributes at a low level ($\sim$ 4\% of the  
DE~\cite{seh92}) to the $\kppee$ branching ratio only.

One item of particular interest is the form factor for the 
DE decay. Historically~\cite{lin88}~\cite{ram93} this has been 
expressed in the form of a $\rho$--propagator modification to the 
magnetic dipole (M1) DE amplitude:
\begin{equation}
\frac{d\Gamma_{M1}}{dE^*_{\gamma}} \sim
                 \left| {\cal F}{\cal M}_{M1} \right|^2 
\end{equation}
where 
\begin{equation}
{\cal F} = \frac{a_1}{(m^2_{\rho}-m^2_K)+2m_{K}E^*_{\gamma}} + a_2
\end{equation}
The effect of the form factor is to soften the photon energy
spectrum of the M1 DE decay. The parametric ratio $a_1/a_2$ 
can then be related to the DE branching ratio by particular 
chiral models~\cite{lin88}~\cite{don86}. 

The decay $\kppee$ which was first reported in the literature by 
KTeV~\cite{ada98}  provides the opportunity for further study of
the $\kppgs$ vertices through the coupling of the $e^+e^-$ decay
plane to the helicity of the photon. Interference between the 
DE and IB photon helicity states results in a manifestly 
$T$--violating angular asymmetry between the $\pi^+\pi^-$ and $e^+e^-$ 
decay planes~\cite{seh92}~\cite{hei93}~\cite{elw95}~\cite{elw96}.

\section{The KTeV Experiment}

\begin{figure}[ht]	% in second brace, h=here, t=top, b=bottom	
\centerline{\epsfxsize 3.0 truein \epsfbox{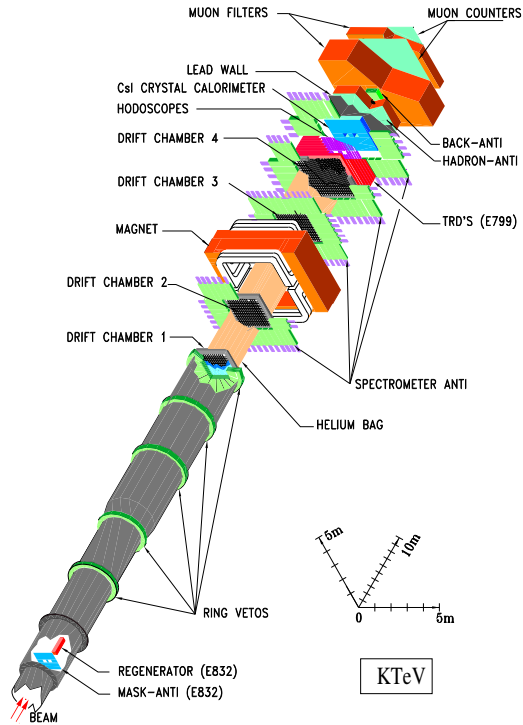}}   
\vskip -.2 cm
\caption[]{
\label{detector}
\small 3D View of the KTeV Detector. }
\end{figure}

KTeV consists of 85 collaborators from 12 institutions from the
U.S. and Japan~\cite{collab}. The primary goal of KTeV is the
measurement of the direct $CP$--violating parameter $\reps$ with
a precision of a part in $10^{-4}$. The detector configuration
is shown in Figure~\ref{detector}. 

The KTeV charged particle detector consists of four drift chambers; 
two upstream and two downstream of a dipole analysing magnet. A
3100 crystal CsI calorimeter provides precise energy measurement of 
photons and pion versus electron particle identification. 

The analyses reported here are from data taken by KTeV between the 
fall of 1996 and the summer of 1997.

\section{$\kppg$ Analysis}

The $\kppg$ events for the analysis presented here were taken
from the fall 1996 KTeV data set. This data accounts for roughly 
a tenth of the $\kppg$ data in hand. Events were selected by
requiring:
\begin{itemize}
\item Two high-quality tracks forming a good vertex and satisfying
      fiducial volume cuts and standard cuts on the veto counters.
      Energy deposited in the calorimeter must not be consistent
      with an electron.
\item At least one additional cluster of energy greater than
      1.5 GeV must be present in the calorimeter.
\item The pion and gamma clusters in the calorimeter must be 
      separated by at least 20 centimeters. This is to supress backgrounds 
      from $\kpip$ decays accompanied by accidental activity or pion 
      hadronic showers.
\item The gamma energy in the center of mass must be greater
      than 20 MeV.
\item The missing $\pi^0$ must have imaginary momentum under the
      $\kpmz$ hypothesis in order to suppress this background.
\end{itemize}
This last requirement is achieved by demanding that the quantity
\begin{equation}
P^2_{\pi^0} \equiv
\frac{(M^2_K - M^2_{\pi^0} - M^2_{\pi\pi})^2 - 4M^2_{\pi^0}M^2_{\pi\pi} 
  - 4M^2_K(P^2_T)_{\pi\pi}}
{4((P^2_T)_{\pi\pi} + M^2_{\pi\pi})}
\end{equation}
be negative ($< - 0.0055$ GeV$^2$).

\begin{figure}[ht]	% in second brace, h=here, t=top, b=bottom	
\centerline{\epsfxsize 3.0 truein \epsfbox{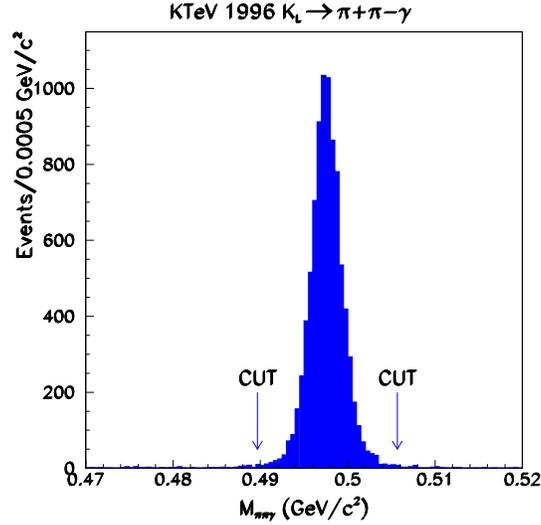}}   
%\vskip -.2 cm
\caption[]{
\label{ppgmass}
\small KTeV 1996 $\pi^+ \pi^- \gamma$ invariant mass distribution, 
       all other cuts have been applied. }
\end{figure}

The resultant very clean $\kppg$ signal is shown in Figure~\ref{ppgmass}.
The remaining background is small (about 0.5\%) and is believed to
consist primarily of residual $\kpip$ decays accompanied by pion
showers in the calorimeter.

\begin{figure}[ht]	% in second brace, h=here, t=top, b=bottom	
\centerline{\epsfxsize 3.0 truein \epsfbox{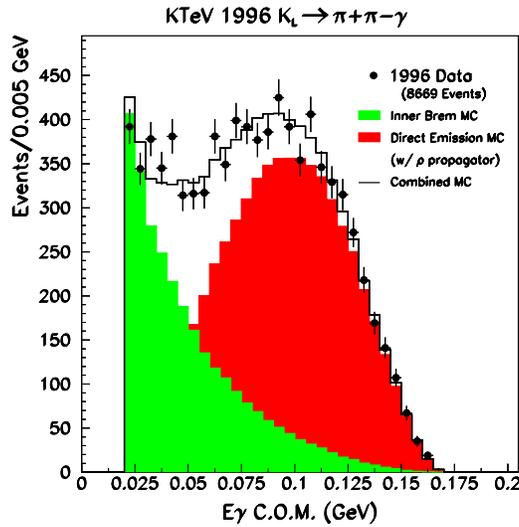}}   
\caption[]{
\label{egcom}
\small KTeV 1996 $\kppg$ photon energy in the center--of--mass.
       Superimposed are the corresponding Monte Carlo IB and 
       ($\rho$--propagator form factor modified) DE decays. MC
       Normalization and $\rho$--propagator parameter $a_1/a_2$ 
       are obtained from Minuit $\chi^2$ minimization. }
\end{figure}

To extract the interesting physics from this decay mode we turn
our attention to the photon energy in the kaon center--of--mass 
system. This distribution for the 1996 data is shown in 
Figure~\ref{egcom} along with the Monte Carlo distributions for
the IB and ($\rho$--propagator form factor modified) DE decays. 

The $\rho$--propagator parameter $a_1/a_2$ and the relative normalization
\begin{equation}
f = \frac{\Gamma(DE)}{\Gamma(DE+IB)}
\end{equation}
are obtained from Minuit $\chi^2$ minimization. The results of the 
fit are $a_1/a_2 = -0.729 \pm 0.026 (stat.)$ and
$\Gamma(DE)/\Gamma(DE+IB) (E^*_{\gamma} > 20 {\rm MeV}) = 
 0.685 \pm 0.009 (stat.)$. Systematic effects in both of these
numbers due to backgrounds and accidental activity are negligible
as are effects due to the variation of most cuts. 2\% systematic
errors are assigned to $a_1/a_2$ due to shifts in the result from
variation of the lower $E^*_{\gamma}$ cut. Systematic errors in the
ratio $\Gamma(DE)/\Gamma(DE+IB)$ are assigned due to shifts in 
the result with variations of the  $E^*_{\gamma}$ cut (2\%) and
$\gamma - \pi$ separation cut (2\%). 

\begin{figure}[ht]	% in second brace, h=here, t=top, b=bottom	
\centerline{\epsfxsize 3.0 truein \epsfbox{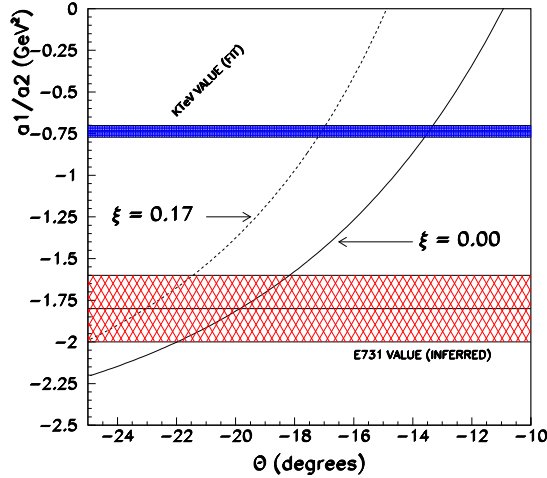}}   
\caption[]{
\label{ampratio}
\small 
  Results of KTeV and previous (FNAL E731) measurements of 
  the $\kppg$ DE form--factor parameter $a_1/a_2$, shown with the 
  theoretical predictions relating $a_1/a_2$ to   
  $\theta \equiv \eta - \eta^{\prime}$ mixing angle and the 
  SU(3) breaking parameter $\xi$~\cite{don86,lin88}.  $\theta$ can
  also be related to the DE branching ratio~\cite{lin88}, which 
  predicts $\theta \sim -20^{\circ} \pm 1^{\circ}$~\cite{ram93}.} 
\end{figure}

Results for the form--factor parameter $a_1/a_2$ differ substantially
from those reported in the literature~\cite{ram93} as illustrated
in Figure~\ref{ampratio}.
The reason for this apparent discrepancy has been traced to the
fact~\cite{ram98} that the E731 result assumed the
Lin and Valencia model in extracting $a_1/a_2$ whereas the KTeV 
result is a true model--independent measurement of this quantity.
Reanalysis of the E731 data by the author indicates that the 
data from the two experiments are completely consistent.

\section{$\kppee$ Analysis}

\begin{figure}[ht]	% in second brace, h=here, t=top, b=bottom	
\centerline{\epsfxsize 3.0 truein \epsfbox{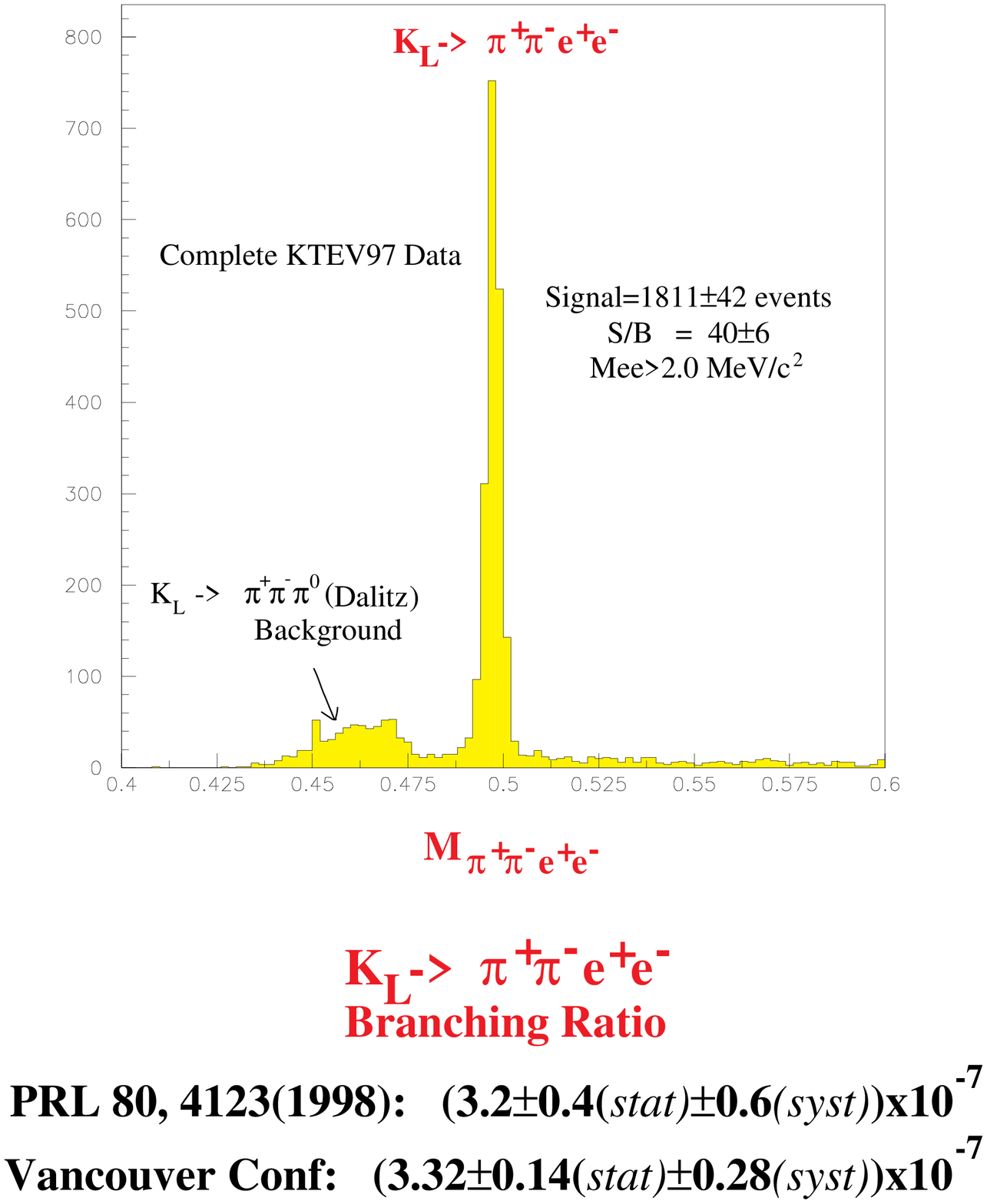}}   
\vspace{0.5cm}
\caption[]{
\label{ppeemass}
\small 
  KTeV $\kppee$ signal and branching ratio.}
\end{figure}

We next turn our attention to the $\kppee$ analysis with 
emphasis on extraction of the $T$--violating angular asymmetry.
KTeV reported the first branching ratio measurement for this
decay~\cite{ada98} based on a subset of the data and has recently
presented an improved number based on the full data set~\cite{bar98}
(See Figure~\ref{ppeemass}). 

As the acceptance of the KTeV spectrometer for the $\kppee$ decay
is a strong function of the virtual photon energy the extraction 
of the angular asymmetry is a two--step process: First the data is
fit to extract the best values of the form factor parameter $a_1/a_2$
and the DE amplitude. Then the Monte Carlo is tuned to the best--fit
values for the final acceptance calculation. 

\begin{figure}[ht]	% in second brace, h=here, t=top, b=bottom	
\centerline{\epsfxsize 3.0 truein \epsfbox{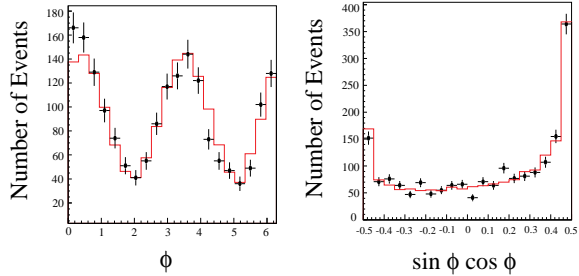}}   
\vspace{0.75cm}
\caption[]{
\label{ppeephi}
\small 
  a) $\phi$ and b) $\sin(\phi)\cos(\phi)$ angular distributions. The
  data are shown as dots, the theoretical expectation (Monte Carlo)
  as a histogram.}     
\end{figure}

Figure~\ref{ppeephi} shows the $\kppee$ angular distributions 
for data and Monte Carlo after the fit for the form factor and
M1 amplitude.  
The raw (before acceptance correction) asymmetry of the data is clear.
After acceptance correction the value of the asymmetry is:
\begin{equation}
\frac{(N_+ - N_-)}{(N_+ + N_-)} = (14.6 \pm 2.3(stat.))\%
\end{equation}
where $N_+$ and $N_-$ are the number of events in the right and left
hand sides --- respectively --- of the $\sin(\phi)\cos(\phi)$ plot
in Figure~\ref{ppeephi}.

Systematic errors of less than a percent are assigned to the asymmetry
from (1) detector resolution (2) uncertainties in the form factor and
M1 amplitude and (3) variations in physics cuts. The total systematic
error on the $\kppee$ plane--angle asymmetry measurement is estimated
to be about 1.1\%.

\section{Summary and Conclusions}

The numerical results presented here are summarized in Table 1.
Here for the first time we present an improved $\kppg$ Direct
Emission branching ratio. We also present the first theory--independent
measurements of a form factor in the $\kppgs$ Direct emission process
in two complimentary decay modes. The observed planar--angle asymmetry
in the $\kppee$ decay is the first instance of $CP$--violation in 
a dynamical variable and is manifestly $T$--violating as well. 
\begin{table}
\caption{Summary of KTeV $\kppg$ and $\kppee$ results.}
\begin{tabular}{ccc} 
                & & \\ 
                & $\kppg$ & $\kppee$ \\ 
                & & \\ 
\tableline 
                & & \\ 
Branching Ratio & $(3.19 \pm 0.09)\times 10^{-5}$ & 
                  $(3.32 \pm 0.14(stat.) \pm 0.28(syst.))\times 10^{-7}$ \\ 
                & (DE Only, $E^*_{\gamma} > 20$ MeV) & \\ 
                & & \\ 
DE/(DE+IB)      & $0.685 \pm 0.009(stat.) \pm 0.017(syst.)$ & --- \\ 
                & & \\ 
$a_1/a_2$       & $-0.729 \pm 0.026(stat.)  \pm 0.015(syst.)$ & 
                  $-0.705^{+0.010}_{-0.020}$ \\ 
                & & \\ 
Asymmetry       & --- & 
                  $(14.6 \pm 2.3(stat.) \pm 1.1(syst.)$\% \\ 
                & & \\ 
\end{tabular}
\end{table}


\begin{references}  % All references should follow standard format

 \bibitem{seh92} Sehgal and Wanniger,  Physical Review {\bf D46},
                 1035 (1992), 5209 (E).

 \bibitem{lin88} Lin and Valencia, Physical Review {\bf D37},
                 143 (1988).  

 \bibitem{ram93} Ramberg {\em et alii}, Physical Review Letters {\bf 70},
                 2525 (1993).  

 \bibitem{don86} Donoghue, Holstein and Lin, Nuclear Physics {\bf B277},
                 651 (1986).  

 \bibitem{ada98} Adams {\em et alii}, Physical Review Letters {\bf 80},
		 4123 (1998).

 \bibitem{hei93} Heiliger and Sehgal, Physical Review {\bf D48},
                 4146 (1993).  

 \bibitem{elw95} Elwood {\em et alii}, Physical Review {\bf D52},
                 5095 (1995), {\bf D53}, 2855 (E,1996).  

 \bibitem{elw96} Elwood {\em et alii}, Physical Review {\bf D53}, 
                 4078 (1996).  

 \bibitem{collab} The KTeV Collaboration: Arizona, UCLA, UCSD, Chicago, 
		  Colorado, Elmhurst, Fermilab, Osaka, Rice, Rutgers, 
		  Virginia, Wisconsin.

 \bibitem{ram98} E. Ramberg, private communication.

 \bibitem{bar98} T. Barker, XXIX International Conference on
		 High--Energy Physics, Vancouver (July 1998).

 \end{references}
\end{document}